\documentclass[twocolumn,tighten]{aastex631}

\usepackage{amsmath}
\usepackage{xspace}
\usepackage{multirow}
\usepackage{fancyapj}
\usepackage{mathtools}
\usepackage{xcolor}

\makeatletter
\def\restartappendixnumbering{\global\applettertrue
\setcounter{table}{0}
\setcounter{figure}{0}
\setcounter{equation}{0}
\def\thetable{\thesection\the\c@table}%
\renewcommand{\theHtable}{Supplement.\thetable}
\def\fnum@table{{\bf\tablename~\thetable}}%
\def\thefigure{\thesection\the\c@figure}%
\def\fnum@figure{{\bf\figurename~\thefigure}}%
}%
\makeatother

  {\list{}{\leftmargin=0.1in\rightmargin=0.1in}\item[]}%
  {\endlist}

\expandafter\def\csname editcolor1\endcsname{magenta}
\expandafter\def\csname editcolor2\endcsname{red}

\newcommand{\E}[1]{\ensuremath{\times 10^{#1}} }

\newcommand{\cts}{\rm\,ct\per{s}\xspace}
\newcommand{\kev}{\rm\,keV\xspace}
\newcommand{\hz}{\rm\,Hz\xspace}
\newcommand{\ks}{\rm\,ks\xspace}

\newcommand{\kpc}{\rm\,kpc\xspace}
\newcommand{\s}{{\rm\,s}\xspace}

\newcommand{\hr}{\rm\,hr\xspace}
\newcommand{\per}[1]{\rm\,#1\ensuremath{^{-1}}\xspace}
\newcommand{\persq}[1]{\rm\,#1\ensuremath{^{-2}}\xspace}
\newcommand{\lumcgs}{\rm\,erg\per{s}\xspace}
\newcommand{\fluxcgs}{{\rm\,erg{\per{s}}{\persq{cm}}\xspace}}
\newcommand{\fluecgs}{\rm\,erg{\persq{cm}}\xspace}

\newcommand{\nicer}{\textrm{NICER}\xspace}

\newcommand{\src}{4U~1730$-$22\xspace}
\newcommand{\srcfull}{4U~1730$-$22\xspace}

\begin{document}
\nolinenumbers

\title{The thermonuclear X-ray bursts of 4U 1730--22}

\author[0000-0002-7252-0991]{Peter Bult}
\affiliation{Department of Astronomy, University of Maryland, College Park, MD 20742, USA}
\affiliation{Astrophysics Science Division, NASA Goddard Space Flight Center, Greenbelt, MD 20771, USA}

\author[0000-0001-9822-6937]{Giulio C. Mancuso}
\affiliation{Instituto Argentino de Radioastronom\'{\i}a (CCT-La Plata, CONICET; CICPBA), C.C. No. 5, 1894 Villa Elisa, Argentina}
\affiliation{Facultad de Ciencias Astron\'omicas y Geof\'{\i}sicas, Universidad Nacional de La Plata, Paseo del Bosque s/n, 1900 La Plata, Argentina}

\author[0000-0001-7681-5845]{Tod E. Strohmayer}
\affil{Astrophysics Science Division and Joint Space-Science Institute, NASA's Goddard Space Flight Center, Greenbelt, MD 20771, USA}

\author[0000-0001-5472-0554]{Arianna C. Albayati}
\affiliation{Physics \& Astronomy, University of Southampton, Southampton, Hampshire SO17 1BJ, UK}

\author[0000-0002-3422-0074]{Diego Altamirano}
\affiliation{Physics \& Astronomy, University of Southampton, Southampton, Hampshire SO17 1BJ, UK}

\author[0000-0002-5341-6929]{Douglas J. K. Buisson}
\affiliation{Independent}

\author[0000-0002-4397-8370]{J\'er\^ome Chenevez}
\affiliation{DTU Space, Technical University of Denmark, Elektrovej 327-328, DK-2800 Lyngby, Denmark}

\author[0000-0002-6449-106X]{Sebastien~Guillot}
\affil{Institut de Recherche en Astrophysique et Plan\'{e}tologie, UPS-OMP, CNRS, CNES, 9 avenue du Colonel Roche, BP 44346, F-31028 Toulouse Cedex 4, France}

\author[0000-0002-3531-9842]{Tolga G\"uver}
\affiliation{Istanbul University, Science Faculty, Department of Astronomy and Space Sciences, Beyaz\i t, 34119, Istanbul, Turkey}
\affiliation{Istanbul University Observatory Research and Application Center, Istanbul University 34119, Istanbul Turkey}

\author[0000-0002-0207-9010]{Wataru Iwakiri}
\affiliation{Department of Physics, Faculty of Science and Engineering, Chuo University, 1-13-27 Kasuga, Bunkyo-ku, Tokyo 112-8551, Japan}

\author[0000-0002-6789-2723]{Gaurava K. Jaisawal} 
\affiliation{DTU Space, Technical University of Denmark, Elektrovej 327-328, DK-2800 Lyngby, Denmark}

\author[0000-0002-0940-6563]{Mason Ng}
\affiliation{MIT Kavli Institute for Astrophysics and Space Research, Massachusetts Institute of Technology, Cambridge, MA 02139, USA}

\author[0000-0002-0118-2649]{Andrea Sanna}
\affiliation{Dipartimento di Fisica, Universit\`a degli Studi di Cagliari, SP Monserrato-Sestu km 0.7, 09042 Monserrato, Italy}

\author[0000-0001-7079-9338]{Jean H. Swank}
\affiliation{Astrophysics Science Division, NASA Goddard Space Flight Center, Greenbelt, MD 20771, USA}

\begin{abstract}\nolinenumbers
  We present observations of the historic transient 4U 1730$-$22 as observed with
  the Neutron Star Interior Composition Explorer (NICER). After remaining in
  quiescence since its 1972 discovery, this X-ray binary showed renewed
  outburst activity in 2021 and 2022. We observed 4U 1730$-$22 extensively with
  NICER, detecting a total of 17 thermonuclear X-ray bursts. From
  a spectroscopic analysis, we find that these X-ray bursts can be divided into
  a group of bright and weak bursts. All bright bursts showed $1\sim2$ second
  rise times and a photospheric radius expansion phase, while the weak bursts showed
  a slower $\sim5$ second rise with a tendency for concave shapes. From the photospheric 
  radius expansion flux, we estimate the source distance at $6.9\pm0.2\kpc$.
  We consider various interpretations for our observations and suggest that
  they may be explained if accreted material is burning stably at the stellar
  equator, and unstable ignition occurs at a range of higher latitudes.
\end{abstract}


\section{Introduction}
\label{sec:intro}
\nolinenumbers

The low-mass X-ray binary \src is a historic transient that was 
discovered in $1972$ with the Uhuru satellite \citep{Cominsky1978}. After an
outburst that lasted about 230 days \citep{Chen1997}, the source returned to
quiescence and has remained dormant for the subsequent $49$ years. Although
\src has been frequently catalogued as a possible neutron star X-ray binary
\citep{Paradijs1995, Chen1997}, the first concrete evidence for this
classification was presented by \citet{Tomsick2007}.  These authors identified
the X-ray source CXOU J173357.5$-$220156 as a candidate quiescent X-ray
counterpart to \src using Chandra observations, finding that the quiescent
spectrum was well described using a neutron star atmosphere model. Given a lack
of further activity from this source, the association between \src and the
quiescent counterpart could not be confirmed. Beyond it being a candidate
neutron star, very little was known about this source. 

Renewed activity in the direction of \src was detected with MAXI/GSC on $2021$
June $7$ \citep{AtelKobayashi21} and initially attributed to a new transient,
tentatively dubbed MAXI~J1733$-$222. Subsequent follow-up observations with
Swift/XRT provided an improved source localization and demonstrated that this
outburst activity was in fact associated with \src \citep{AtelKennea21a,
AtelKennea21b}. This association was later confirmed by the detection of an
optical counterpart \citep{ATelRussell21, AtelStrader21}, whose precise
location matches both the Swift localization and the quiescent source position
of \citet{Tomsick2007}.

After maintaining a relatively low X-ray flux of $\sim10^{-10}\fluxcgs$ for
a few weeks\citep{AtelKennea21a}, the source started brightening substantially
on July 5, reaching an order of magnitude increase in X-ray flux by July~6
\citep{ATelIwakari21} as it transitioned from a hard to a soft accretion state.
On July 7, the first thermonuclear (type I) X-ray burst from the source was
detected with NICER \citep{AtelBult21e}, confirming that \src indeed harbors an
accreting neutron star. The intensity of the source gradually decayed over the
subsequent $\sim100$ days, although for a lack of pointed observations it is
unclear if and when \src returned to quiescence.

On 2022 February 13 the MAXI Nova alert system \citep{Negoro2016} again
triggered on \src as the source unexpectedly returned to a bright state. At
this time, we began a regular monitoring campaign with NICER. We found that \src
remained in its bright soft state for about 150 days, only returning to a hard
state in early July 2022.  Over the course of its prolonged soft state phase,
we detected sixteen more X-ray bursts. An independent analysis of these data
recently reported the detection of a 585\hz burst oscillation in one
of the X-ray bursts \citep{Li2022}. 

In this paper we combine the 2021 and 2022 \nicer observations of \src to present a detailed
spectroscopic analysis of all detected X-ray bursts from this source. 

\section{Observations}
We observed \src with \nicer in 2021 and 2022 for a total unfiltered exposure
of $102\ks$ and $574\ks$, respectively. These observations are collected under
ObsIDs starting with $420220$, $463901$, and $520220$. We processed the data
using \textsc{nicerdas} version 9, as distributed with \textsc{heasoft} version
6.30. All standard filter criteria were applied, meaning that we retained only
those epochs during which the pointing offset was $<54\arcsec$, the Earth
elevation angle was $>15\arcdeg$, the elevation angle with respect to the
bright Earth limb was $>30\arcdeg$, and the instrument was not in the South
Atlantic Anomaly (SAA). Additionally, we applied standard background filter
criteria: we removed all epochs during which the rate of detected
reset triggers per detector (undershoots) is larger than $500\cts$ or when the rate of high
energy events per detector (overshoots) is either greater than 1.5 or greater than $1.5
\times \textsc{cor\_sax}^{-0.633}$, where \textsc{cor\_sax} gives the
geomagnetic cut-off rigidity in units of GeV\per{c}. To prevent the overshoot
filters from introducing spurious $1-10$\s gaps in the data, we followed \citet{Bult2020a}
and applied a 5\s window smoothing to the overshoot rates prior
to evaluating the filter condition. Finally, we added the
requirement that all 52 detectors were active during the
observation.

The filter criteria described above yielded clean data products for the vast
majority of ObsIDs analysed in this paper. However, in 21 ObsIDs we found
periods of low level background flaring related to polar horn passages \citep{Remillard2022}
that were not fully removed. For these ObsIDs we removed the epochs during which the cutoff
rigidity (\textsc{cor\_sax}) was smaller than $1.5$\,Gev\per{c}. In all cases,
this appropriately removed the background dominated epochs.

After processing, we were left with $72\ks$ and $430\ks$ clean exposure for the
2021 and 2022 outbursts, respectively. Visually inspecting the light curves of these
data, we identified $16$ thermonuclear (type I) X-ray bursts. Comparing to the
unfiltered light curve, we identified one additional X-ray burst during an SAA
passage. Because the background was only modestly elevated during
the SAA passage in which this additional X-ray burst was detected, we included
this epoch in our analysis. The ObsIDs and occurrence times of these bursts are
reported in Table \ref{tab:burst properties}.

\begin{table*}[t]
  \centering
  \movetableright=-0.75in
  \caption{%
    X-ray burst properties
    \label{tab:burst properties}
  }
  \newcommand{\tbm}[1]{\tablenotemark{#1}}
  \begin{tabular}{l l l l l l l h l l l l}
    \hline \hline
    No &  ObsID              &  MJD           & PRE  & Peak flux            & Fluence               & Rise  &  $\tau$ & $\epsilon_1$ & $\epsilon_2$ & $\alpha$  & t$_{\rm rec}$ \\ 
    ~  &  ~                  &  (TT)          &      & ($\E{-8} \fluxcgs$)  & ($\E{-7} \fluecgs$)   & (s)   &  (s)    & (s)          & (s)          & ~         & (hr)   \\
    \tableline
    1   &  4202200125         &  59404.552433  &  n   & $4.9 \pm 1.1$        & $3.73 \pm 0.08$       &  5.4  &    7.6  &   -   &   8.2  &  $512_{-102}^{+154}$  &  ~  \\
    2   &  5202200101         &  59639.336264  &  n   & $6.5 \pm 1.2$        & $5.39 \pm 0.11$       &  3.6  &    8.2  &   -   &   9.5  &  $339_{- 68}^{+102}$  &  ~  \\
    3   &  5202200112\tbm{a}  &  59657.912770  &  y   & $6.2 \pm 0.6$        &$>5.80 \pm 0.09$       &  1.6  &    9.4  &  2.1  &  16.4  &  $368_{- 74}^{+110}$  &  $>3.8 \pm 0.3$ \\
    4   &  5202200113         &  59658.960817  &  y   & $7.7 \pm 1.5$        & $8.44 \pm 0.10$       &  2.1  &   11.0  &  1.9  &  15.1  &  $251_{- 50}^{+ 75}$  &  $ 5.6 \pm 0.5$ \\
    5   &  4639010102         &  59664.121513  &  y   & $7.7 \pm 1.4$        & $9.58 \pm 0.11$       &  1.6  &   12.4  &  2.0  &  14.3  &  $211_{- 42}^{+ 63}$  &  $ 6.7 \pm 0.6$ \\
    6   &  4639010104         &  59666.949981  &  n   & $5.2 \pm 0.9$        & $3.85 \pm 0.17$       &  4.6  &    7.4  &   -   &   7.2  &  $585_{-117}^{+175}$  &  ~  \\
    7   &  4639010113         &  59675.595021  &  n   & $2.8 \pm 0.4$        & $2.92 \pm 0.03$       &  2.1  &   10.3  &   -   &   8.4  &  $686_{-137}^{+206}$  &  ~  \\
    8   &  4639010116         &  59678.770816  &  y   & $7.0 \pm 0.5$        & $7.94 \pm 0.10$       &  1.6  &   11.4  &  2.3  &  14.9  &  $227_{- 45}^{+ 68}$  &  $ 6.2 \pm 0.5$ \\
    9   &  4639010131         &  59695.093117  &  y   & $7.5 \pm 0.7$        & $6.79 \pm 0.08$       &  1.1  &    9.1  &  2.1  &  13.5  &  $230_{- 46}^{+ 69}$  &  $ 6.1 \pm 0.5$ \\
    10  &  4639010141         &  59718.325513  &  y   & $6.2 \pm 0.5$        & $7.63 \pm 0.10$       &  1.1  &   12.3  &  3.4  &  13.7  &  $200_{- 40}^{+ 60}$  &  $ 7.1 \pm 0.6$ \\
    11  &  4639010146         &  59723.819599  &  n   & $4.5 \pm 0.9$        & $3.73 \pm 0.07$       &  5.6  &    8.2  &   -   &   9.0  &  $457_{- 91}^{+137}$  &  ~  \\
    12  &  4639010160         &  59739.423071  &  y   & $7.0 \pm 1.0$        & $6.89 \pm 0.08$       &  1.1  &    9.9  &  2.8  &  13.4  &  $189_{- 38}^{+ 57}$  &  $ 7.5 \pm 0.6$ \\
    13  &  4639010160         &  59739.868686  &  y   & $8.1 \pm 1.0$        & $8.22 \pm 0.09$       &  1.1  &   10.1  &  2.1  &  13.9  &  $132_{- 26}^{+ 39}$  &  $10.7 \pm 0.9$ \\
    14  &  4639010166         &  59747.677906  &  y   & $6.7 \pm 0.4$        & $9.26 \pm 0.11$       &  1.6  &   13.9  &  2.3  &  14.8  &  $162_{- 32}^{+ 49}$  &  $ 8.7 \pm 0.7$ \\
    15  &  4639010175         &  59756.854948  &  y   & $6.5 \pm 1.5$        & $9.04 \pm 0.09$       &  1.9  &   14.0  &  2.9  &  14.9  &  $100_{- 20}^{+ 30}$  &  $14.0 \pm 1.2$ \\
    16  &  4639010177\tbm{a,b}&  59760.004202  &  y   & $5.3 \pm 0.6$        &$>5.20 \pm 0.06$       &  1.4  &    9.8  &  4.4  &  10.0  &  $194_{- 39}^{+ 58}$  &  $>7.3 \pm 0.6$ \\
    17  &  4639010179         &  59762.731650  &  y   & $7.1 \pm 1.3$        & $8.89 \pm 0.24$       &  2.4  &   12.5  &  2.4  &  17.2  &  $ 98_{- 20}^{+ 29}$  &  $14.4 \pm 1.2$ \\
    \tableline
  \end{tabular}
  \flushleft
  \tablenotetext{a}{Truncated by the end of the observation.}
  \tablenotetext{b}{Burst detected during SAA.}
  \tablecomments{The MJD column lists the onset time of the burst. The ``PRE'' column
  indicates if the burst exhibited photospheric radius expansion (y) or not (n). 
  The peak flux and fluence are both bolometric. 
  The ``Rise'' column gives the rise time of the bursts, while $\epsilon_1$ and $\epsilon_2$ give the e-folding timescales (see Section \ref{sec:light curves}).
  The $\alpha$ column (see Eq. \ref{eq:alpha}) was calculated assuming a burst recurrence time of $10$ hours,
  whereas the t$_{\rm rec}$ column gives the recurrence time if we instead assume $\alpha=140$
  (see Section \ref{sec:burst interpretation} for a discussion of these columns).
  Uncertainties are quoted at $68\%$ confidence.}
\end{table*}

\section{Results}

\subsection{Light curves}
\label{sec:light curves}

\begin{figure*}[t]
  \includegraphics[width=\linewidth]{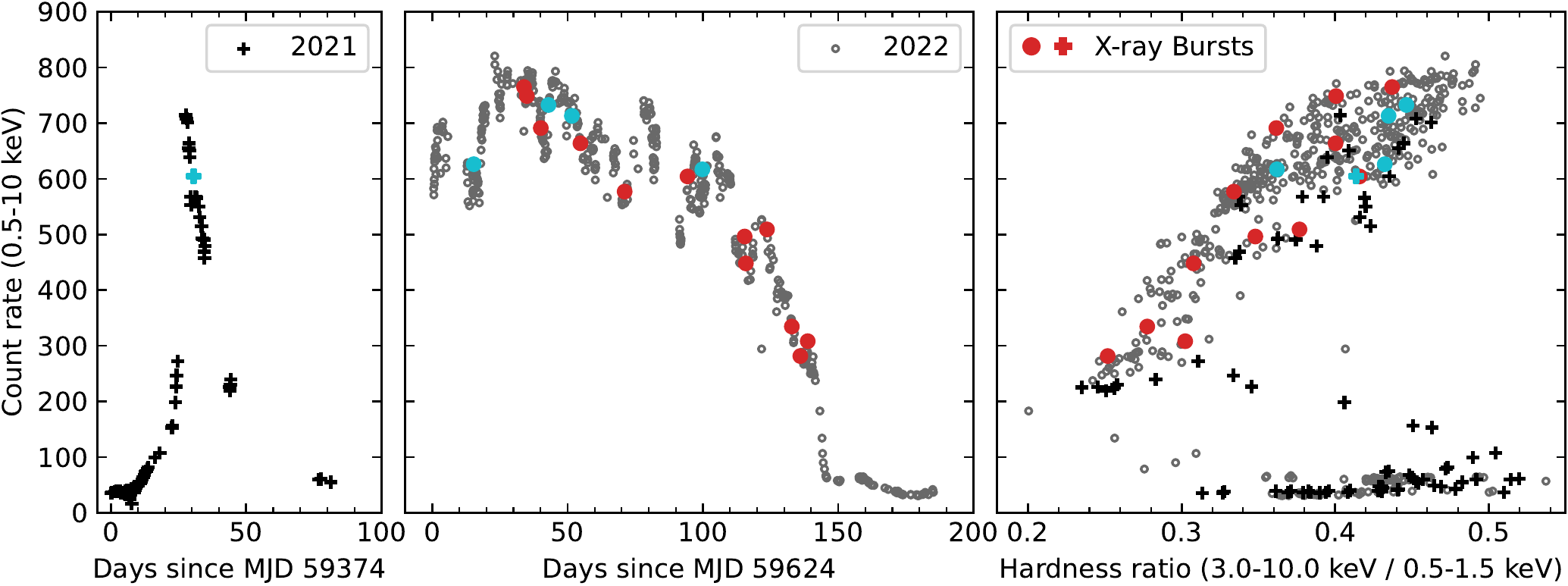}
  \caption{%
    Light curves in the $0.5-10\kev$ band (left, middle) and hardness intensity diagram
    (right) of \src, with each point representing a single \nicer pointing. Those
    pointings containing an X-ray burst are marked in teal (weak) and red
    (bright). 
  } 
  \label{fig:light curve}
\end{figure*}

We group the data by continuous pointing. We find there are $607$ such
pointings in the dataset, with exposures ranging from $150\s$ to $2500\s$.  For
each pointing we calculate the average count-rate in the $0.5-10\kev$ energy
band, as well as a hardness ratio defined as the $3-10\kev$ rate over the
$0.5-1.5\kev$ rate. The resulting light curve and hardness evolution are shown
in Figure \ref{fig:light curve}. 
We see that our observations sample the source in two states: a fainter hard
state at count-rates of $\approx 50\cts$ and a brighter soft state with
count-rates $\gtrsim250\cts$.  All X-ray bursts were detected in the soft
state. 

\begin{figure*}[t]
  \includegraphics[width=\linewidth]{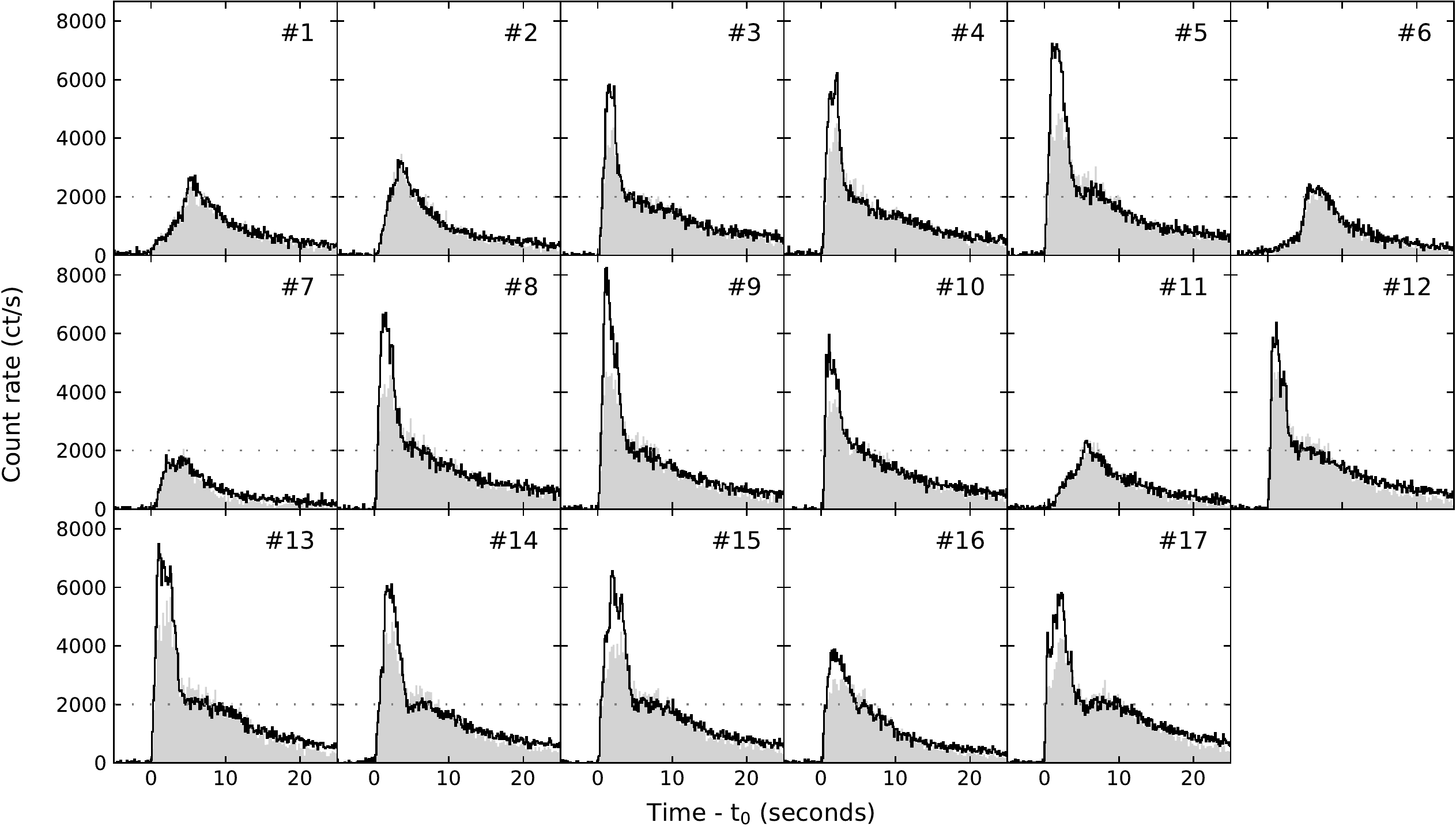}
  \includegraphics[width=\linewidth]{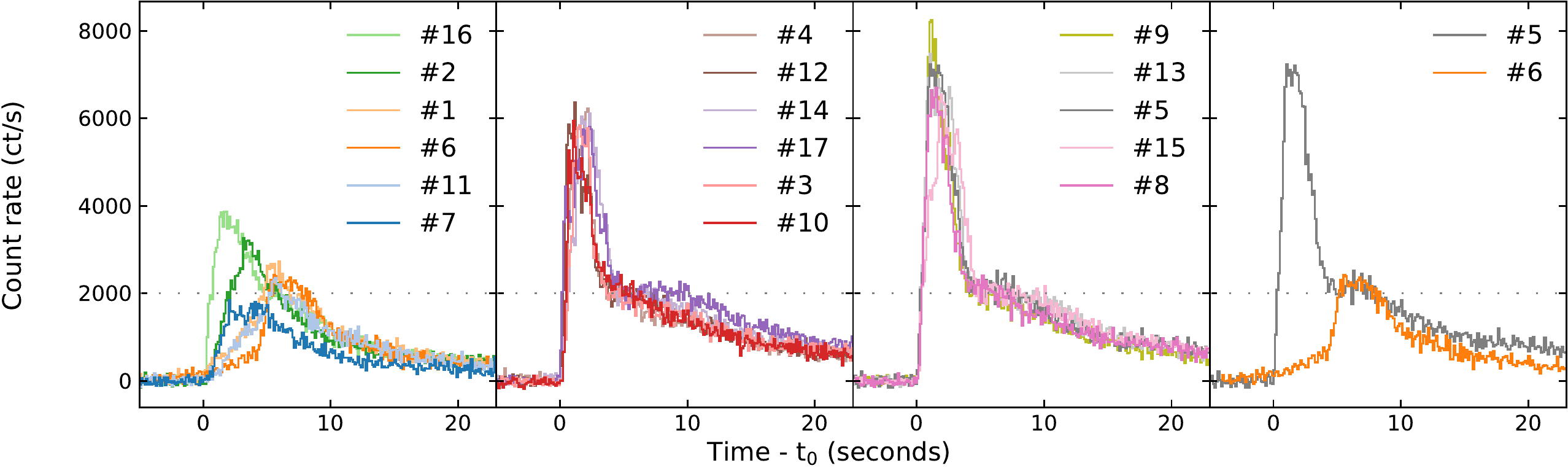}
  \caption{%
    Light curve profiles of the 17 X-ray bursts observed from \src. The
    top panels show the individual bursts as labeled, with the black line
    showing the $0.5-10\kev$ count-rate at $1/8\s$ time resolution and the gray
    area showing the $4-10\kev$ count-rate ($1/4\s$ resolution, multiplied by
    a factor four). The four panels in the bottom row show various groupings of
    bursts for comparison, with both labels and color coding ordered by
    increasing peak count-rate. All curves are plotted relative to their
    respective burst onset time (Table \ref{tab:burst properties}) and with the pre-burst
    count-rate subtracted.  Additionally, a horizontal dotted line was added to each panel 
    at $2000$\cts to guide the eye.
  } 
  \label{fig:burst lc}
\end{figure*}

The X-ray burst light curves show most of their flux during the first $\approx20\s$,
but show long low intensity tails, taking about $\approx100\s$ for the burst
count-rate to return to the pre-burst level. In spite of the long tail, we
observed the entire X-ray burst in all cases except bursts $\#3$ and $\#16$.
For these two bursts the tail was truncated by the end of the observation at
$>30\s$ after onset. In Figure \ref{fig:burst lc} we show the light curves of
the $17$ observed X-ray bursts in the $0.5-10\kev$ band (black line) and
$4-10\kev$ band (gray area). Based on the morphology of their pre-burst rate
subtracted profiles, we can divide the bursts into two categories: a weak and
a bright group. 

The group of weak bursts is characterized by slow rise times of about
$5\s$ and peak count-rates on the order of $2000\cts$. 
The group of bright bursts show a more rapid rise, taking about $1-2\s$
to reach peak count-rates between $6000-8000\cts$. These brighter bursts then
consistently show an initial sharp decline in count-rate back to $2000\cts$.
From that point on the burst decay transitions into a slower trend that is
similar to the decays seen in the weak bursts. Further, we note that a number
of the bright bursts show a temporary plateau at $2000\cts$ before continuing
their decay. See, for instance, bursts $\#5$ and $\#17$ in Figure
\ref{fig:burst lc}. The initial fast rise and rapid decay of the bright group
suggests that the ignition of these bursts occurs in a hydrogen poor environment,
while the much slower rise of the weak group instead points to ignition in
a hydrogen rich environment \citep{Galloway2021}.

It is clear that for both weak and bright bursts the $2000\cts$ rate
signifies some special state in the burst evolution. Another aspect of this
behavior is that the inflection in the tail of the bright bursts
consistently occurs about five seconds after onset. This aligns with the time
it takes for the weak bursts to reach their peak intensity. This phenomenon
is illustrated in the bottom right panel of Figure \ref{fig:burst lc}, where
we plot bursts $\#5$ and $\#6$ together.

Looking closer at the group of bright bursts, we also note that the initial
peak is less pronounced in the $4-10$ keV light curves. Hence, this peak
appears to be predominantly driven by softer photons, suggesting the presence
of a photospheric radius expansion phase. Further, by comparing the different
burst profiles (bottom row of Figure \ref{fig:burst lc}), we see that while
many of these bursts have generally similar peak rates, some bursts are notably
brighter (\#5, \#9, \#13, and \#15), while \#2 and \#16 fall somewhere in
between the two groups. 

To quantify the burst shapes, we define the burst rise time as the time it
takes from the onset to reach peak intensity, with the onset determined
visually to optimize the alignment between the bursts.  For each burst we also attempted
to measure the exponential decay timescale. Because an exponential function is
unable to account for the break observed in the bright bursts, we split the
burst tails into phases. For each of the bright bursts we measure the
exponential decay timescale, $\epsilon_1$, between $[t_0 + 2\s, t_0 + 5\s]$,
where $t_0$ is the respective burst onset time. For both the faint and bright bursts
we measure a second decay timescale between $[t_0+5\s, t_0+30\s]$, which we
call $\epsilon_2$. The rise times and exponential decay timescales are reported
in Table \ref{tab:burst properties}.

\subsection{Time-resolved spectroscopy}

We perform a spectroscopic analysis of each of the observed X-ray bursts using
\textsc{xspec} v12.12.1c \citep{Arnaud1996}. The interstellar absorption is
modeled using the T\"ubingen-Boulder model \citep{Wilms2000}. Background
spectra were generated using the \nicer 3C50 model \citep{Remillard2022}.
All X-ray spectra were binned used the optimal binning method of \citep{Kaastra2016},
additionally requiring at least 25 events per spectral channel. 

We begin by considering the spectrum of the persistent (non-burst) emission
around the time each X-ray burst was observed. We extracted a pre-burst spectrum
in the $0.5-10\kev$ range from the epoch $[t_0-225\s, t_0-25\s]$, where $t_0$
refers to the respective burst onset time. We model these pre-burst spectra
using an absorbed multi-temperature disk blackbody (\texttt{diskbb},
\citealt{Mitsuda1984, makishima1986}) plus a thermally Comptonized continuum
(\texttt{nthcomp}, \citealt{Zdziarski1996, Zycki1999}), which yields
a reasonably good description of the continuum emission in all cases. 
We further use the \texttt{cflux} model component to estimate the unabsorbed
bolometric flux by extrapolating our model over the $0.01-100$\kev energy range.
We find a consistent absorption column density across all pre-burst spectra,
with a mean value of $N_{\rm H} = (7.1\pm0.3) \E{21} \persq{cm}$.  Considering the
pre-burst spectra as a function of flux, we observe an evolution in the disk
component, which increases in temperature from about 0.8\kev to $1.8\kev$.  The
photon index remains constant in all spectra, with an average value of $2.0 \pm
0.1$, while the Comptonization normalization increases from $0.2$ to $0.4$
photons\per{keV}\persq{cm}\per{s} at 1\kev. The detailed best-fit parameters
are listed in Table \ref{tab:preburst}.

\begin{table*}[t]
  \centering
  \movetableright=-0.75in
  \caption{%
    Preburst spectroscopy
    \label{tab:preburst}
  }
  \begin{tabular}{l l l l l l l l}
    \hline \hline
    No &  $N_{\rm H}$              & T$_{\rm in}$ (keV)       & diskbb norm            & Photon Index             & nthcomp norm             & Flux                       & $\chi^2$/dof    \\
     ~  &  ($\E{21}\persq{cm}$)    & (keV)                    & ~                      & ~                        & ~                        & ($\E{-9}\fluxcgs$)         & \\
    \tableline
     1  &  $6.9\pm0.2$            & $1.48_{-0.10}^{+0.09}$   & $12.0_{-2.0}^{+3.5}$   & $1.93\pm0.10$            & $0.30\pm0.03$            & $5.3\pm0.2$             & 139.1/113       \\
     2  &  $7.1\pm0.2$            & $1.74_{-0.12}^{+0.10}$   & $7.5_{-1.1}^{+1.5}$    & $2.05_{-0.13}^{+0.17}$   & $0.34\pm0.03$            & $5.1_{-0.3}^{+0.6}   $ & 120.6/108       \\
     3  &  $7.2\pm0.2$            & $1.57_{-0.09}^{+0.10}$   & $13_{-2}^{+3}$         & $1.94_{-0.08}^{+0.12}$   & $0.41\pm0.03$            & $5.9_{-0.2}^{+0.5}   $ & 157.0/117       \\
     4  &  $7.1\pm0.2$            & $1.37_{-0.11}^{+0.10}$   & $18_{-3}^{+7}$         & $1.95\pm0.10$            & $0.41_{-0.05}^{+0.04}$   & $5.9_{-0.3}^{+0.5}   $ & 170.9/114       \\
     5  &  $7.4\pm0.2$            & $1.23_{-0.08}^{+0.10}$   & $21_{-5}^{+6}$         & $2.00_{-0.06}^{+0.08}$   & $0.43_{-0.04}^{+0.03}$   & $5.6_{-0.3}^{+0.4}   $ & 158.5/112       \\
     6  &  $7.4\pm0.2$            & $1.79_{-0.11}^{+0.07}$   & $8.8_{-1.0}^{+1.3}$    & $2.12_{-0.15}^{+0.19}$   & $0.40\pm0.03$            & $6.2_{-0.6}^{+1.0}   $ & 171.8/115       \\
     7  &  $7.1\pm0.2$            & $1.61_{-0.14}^{+0.13}$   & $10.0_{-1.8}^{+3.5}$   & $1.96_{-0.10}^{+0.14}$   & $0.39_{-0.03}^{+0.02}$   & $5.6\pm0.4            $ & 166.1/115       \\
     8  &  $7.1\pm0.2$            & $1.34_{-0.09}^{+0.10}$   & $16_{-4}^{+5}$         & $1.92\pm0.08$            & $0.37\pm0.04$            & $5.0_{-0.2}^{+0.7}   $ & 138.8/114       \\
     9  &  $7.1\pm0.3$            & $1.10_{-0.08}^{+0.09}$   & $28_{-8}^{+11}$        & $1.96_{-0.10}^{+0.09}$   & $0.33\pm0.04$            & $4.3\pm0.3            $ & 135.3/110       \\
    10  &  $6.9\pm0.3$            & $1.26_{-0.09}^{+0.10}$   & $18_{-5}^{+7}$         & $1.81_{-0.10}^{+0.09}$   & $0.30\pm0.04$            & $4.2_{-0.2}^{+0.3}   $ & 153.5/114       \\
    11  &  $7.0\pm0.2$            & $1.31_{-0.08}^{+0.12}$   & $17\pm4$               & $1.95_{-0.07}^{+0.10}$   & $0.36\pm0.03$            & $4.7_{-0.4}^{+0.5}   $ &  95.2/112       \\
    12  &  $7.0\pm0.3$            & $1.08_{-0.06}^{+0.11}$   & $24\pm8$               & $1.86_{-0.08}^{+0.10}$   & $0.28\pm0.04$            & $3.6_{-0.3}^{+0.4}   $ & 182.6/109       \\
    13  &  $6.7\pm0.3$            & $0.88_{-0.04}^{+0.05}$   & $54_{-16}^{+17}$       & $1.81_{-0.13}^{+0.12}$   & $0.21_{-0.04}^{+0.05}$   & $3.00_{-0.15}^{+0.34}$ & 116.5/108       \\
    14  &  $7.2\pm0.2$            & $1.33_{-0.12}^{+0.15}$   & $10_{-3}^{+4}$         & $1.95_{-0.07}^{+0.09}$   & $0.33\pm0.03$            & $4.2_{-0.2}^{+0.4}   $ & 170.4/111       \\
    15  &  $7.0\pm0.3$            & $0.78\pm0.06$            & $48_{-18}^{+25}$       & $1.93_{-0.12}^{+0.10}$   & $0.21\pm0.04$            & $2.52_{-0.10}^{+0.43}$ & 109.8/102       \\
    16  &  $7.4_{-0.4}^{+0.3}$   & $0.82\pm0.09$            & $28_{-12}^{+21}$       & $2.16_{-0.12}^{+0.09}$   & $0.23_{-0.04}^{+0.03}$   & $2.8_{-0.3}^{+0.4}   $ & 133.3/ 99       \\
    17  &  $7.1_{-0.4}^{+0.3}$   & $0.80_{-0.11}^{+0.14}$   & $31_{-16}^{+31}$       & $1.89_{-0.11}^{+0.09}$   & $0.21\pm0.03$            & $2.4\pm0.3            $ &  97.0/104       \\
    \tableline
  \end{tabular}
  \flushleft
  \tablecomments{We report the unabsorbed bolometric flux. Uncertainties are quoted at $90\%$ confidence. 
  }
\end{table*}

To analyze the emission of the X-ray bursts themselves, we extracted the burst
epochs between $[t_0-5\s ,t_0+95\s]$, using all events in the $0.5-10\kev$
energy range. We then proceeded to dynamically bin the X-ray burst data
such that each bin contained $2000$ events. The time-resolved burst spectra were
extracted from these dynamic bins.

We initially modeled the burst spectra using a simple absorbed blackbody model. That
is, we added a blackbody component to the respective pre-burst model, and fixed
all model parameters but the blackbody normalization and temperature.
This approach yielded a very poor description of the data. The best-fit reduced
$\chi^2$ scores were found to increase with the source count-rate, peaking at
scores of $3-5$ (with associated p-values of $<10^{-13}$) around the times of
peak burst count-rates. 

\begin{figure}[t]
  \includegraphics[width=\linewidth]{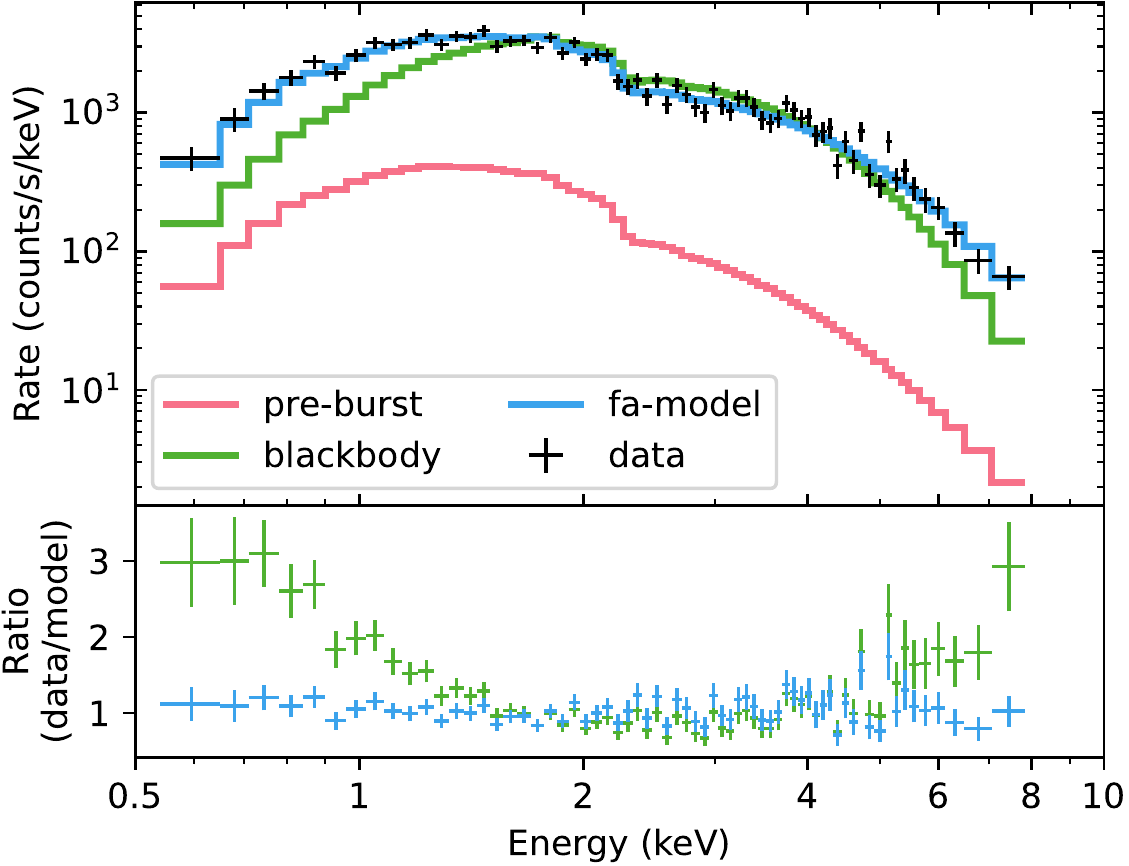}
  \caption{%
    Best-fit spectra of burst \#5 at peak intensity, showing in the top panel:
    the measured spectrum (black), the pre-burst model (red), a blackbody model (green),
    and the fa-model (blue). The bottom panel gives the ratio of the data over the
    respective models. Note that the spectral bin above $8\kev$ had an insufficient event
    count and was removed from the fit. 
  } 
  \label{fig:spectrum}
\end{figure}

\begin{figure*}[t]
  \centering
  \includegraphics[width=0.49\linewidth]{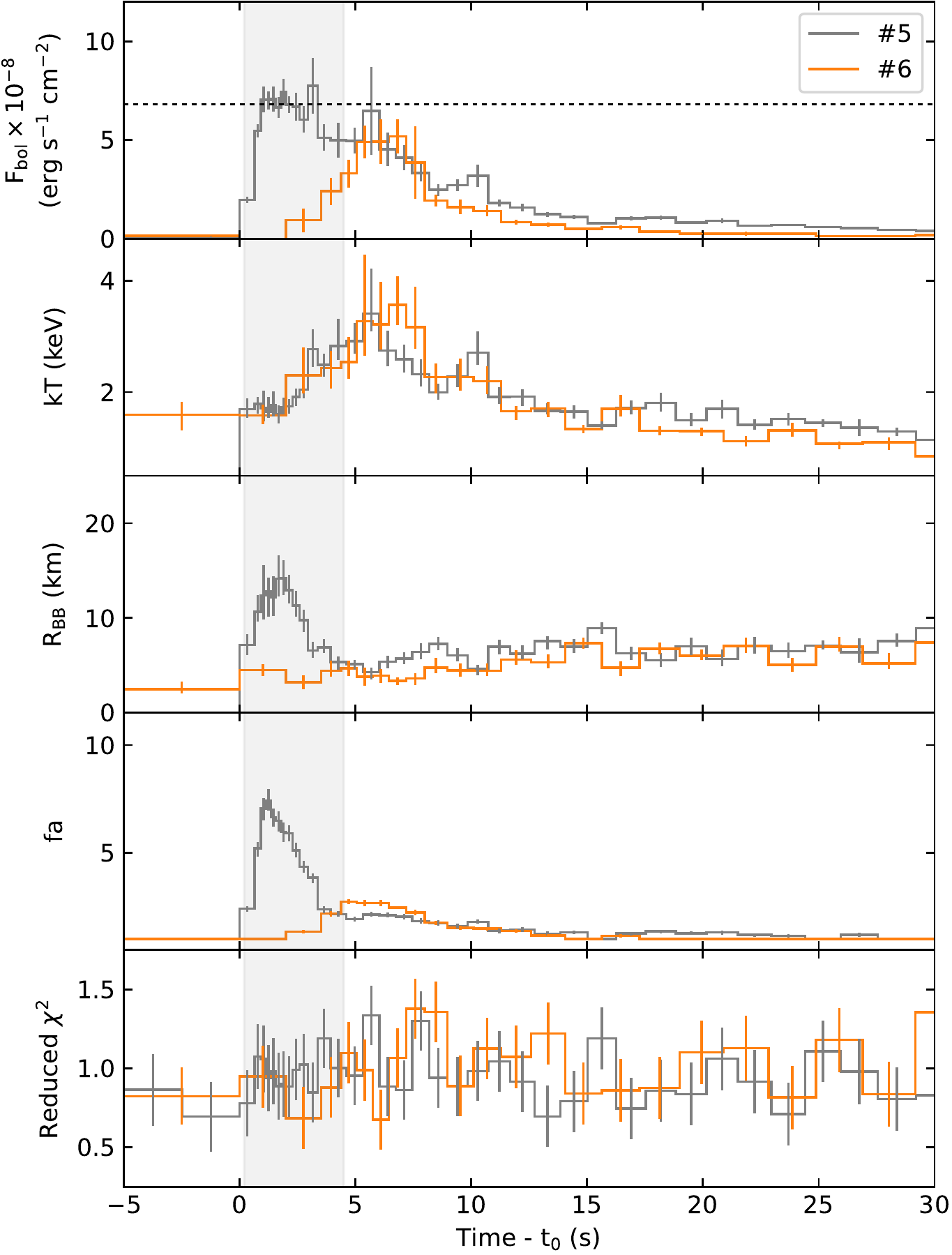}
  \includegraphics[width=0.49\linewidth]{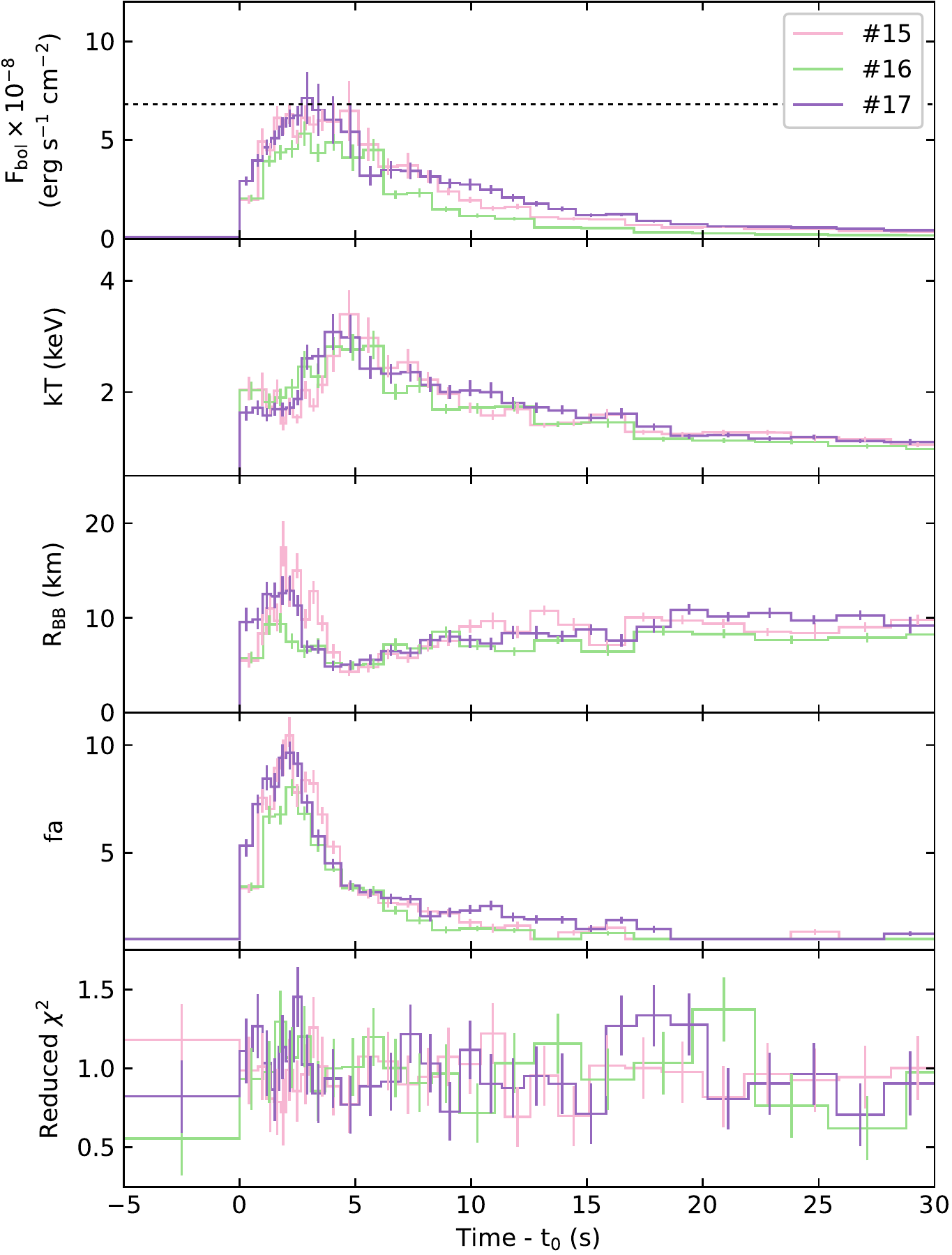}
  \caption{%
    Time resolved spectroscopy of bursts $\#5$ (left, gray), $\#6$ (left, orange), 
    $\#15$ (right, pink), $\#16$ (right, green), and $\#17$ (right, purple),
    using the same color coding as in Figure \ref{fig:burst lc}. Both figures
    show from top to bottom: the bolometric burst flux, the blackbody
    temperature, the blackbody radius (at 6.9\kpc), the $f_a$ factor, and the reduced
    $\chi^2$. We further highlighted the epoch of radius expansion for burst \#5
    in the left panel. The error bars show $1\sigma$ uncertainties, which for the bottom
    panels were calculated as $\sqrt{2/\mbox{dof}}$, with dof the degrees of
    freedom. The dashed line in the top panels indicates the photospheric
    radius expansion flux. 
  } 
  \label{fig:trs}
\end{figure*}

In a second approach we rescaled the pre-burst model component using a variable
factor, $f_a$ \citep{Worpel2013, Worpel2015}. This approach greatly improved
the model fits, yielding acceptable $\chi^2$ scores throughout each of the
bursts. In Figure \ref{fig:spectrum} we show an example spectrum comparing the blackbody
model to the $f_a$ model fit, while in Figure \ref{fig:trs} we show the resulting time-resolved
spectroscopic evolution for a few example bursts. We find that measured $f_a$
values roughly follow the burst light curves, peaking at values of about $3$
for the group of weak bursts, while the group of bright bursts show $f_a$
values as high as $10$. Key burst parameters are listed in Table \ref{tab:burst
properties}. 

Each of the X-ray bursts in the bright group shows photospheric radius
expansion (PRE): the blackbody temperature dips and then rises, while simultaneously
the blackbody normalization peaks and then decays. Notably, we find that
``touchdown'' (the time at which the blackbody peaks in temperature) occurs at
$t_0 + 5\s$ - the same time at which these bright bursts show an inflection in
their light curves. 

For each modeled burst spectrum we calculate the bolometric flux contributed
by the blackbody. We further estimate the burst flux that is not captured by
the blackbody as $(f_{a}-1) \times F_{\rm preburst}$, where $F_{\rm preburst}$ is the
bolometric flux measured for the respective pre-burst spectrum (Table
\ref{tab:preburst}). Finally, we define the total burst flux as the sum of the
blackbody and secondary emission components. 

Considering the evolution of the flux over time, we find that the
blackbody flux systematically drops during the PRE phase, while the total burst flux
remains constant in time. This pattern is illustrated in Figure \ref{fig:flux
temp}, where we show the bolometric flux of these two components as a function
of blackbody temperature. 
The upper branch of the track represents the PRE phase, moving toward the right
as the photosphere expands in radius. The flux contributed by the blackbody (black points)
decreases with radius, while the total burst flux (red points) remains constant.
We therefore adopt the total burst flux as the more accurate measure for the burst
energy. To estimate the PRE flux, we average all burst flux measurements during
the PRE phase of the bursts that show one, which gives us $F_{\rm PRE}
= (6.75\pm0.25)\E{-8} \fluxcgs$. 

\begin{figure}[t]
\includegraphics[width=\linewidth]{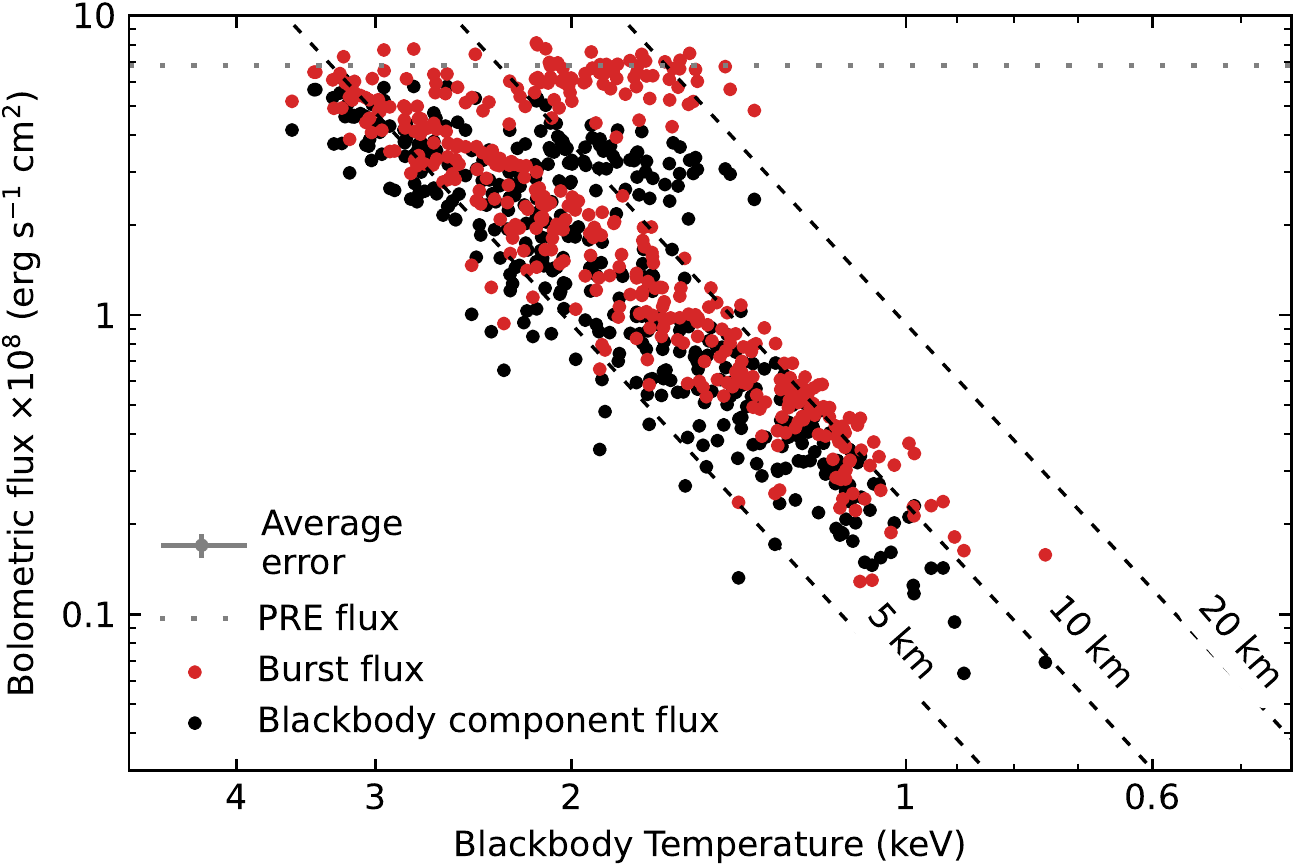}
\caption{%
  Flux temperature diagram for the X-ray bursts of \src. The bolometric flux
  contributed by the blackbody component is shown in black, while the red
  points show the bolometric flux of the whole burst spectrum (blackbody plus
  the excess, but not including the pre-burst contribution; see text). The
  horizontal dotted line indicates the photospheric radius expansion flux,
  while the three diagonal dashed lines show the constant radius contours as
  labeled, assuming a source distance of $6.9\kpc$. These contours are for
  illustrative purposes only, as they do not take color corrections into
  account \citep[see, e.g.,][]{Suleimanov2017}. Error bars were omitted
  for clarity, but see the grey point in the legend for the average scale
  of the uncertainty. 
} 
\label{fig:flux temp}
\end{figure}

\section{Discussion}
We have presented a spectroscopic analysis of $17$ thermonuclear X-ray bursts
observed from \src with \nicer. We found that the burst light curves can be
divided into two groups: slow rising weak bursts and fast rising bright
bursts. We observed photospheric radius expansion in each of the bright bursts,
and estimated the bolometric burst flux during the PRE phase to be $F_{\rm PRE}
= (6.75\pm0.25)\E{-8}\fluxcgs$. Equating this flux to the $3.8\E{38}\lumcgs$ empirical
Eddington luminosity of \citet{Kuulkers2003}, we obtain a source distance
estimate of $6.9\pm0.2\kpc$. 
  An important caveat to this distance estimate is that we choose to use the
  total burst flux to estimate the PRE flux (see Section \ref{sec:nonthermal}
  for further discussion on this point). If we instead use the blackbody
  contribution only, then the PRE flux is about 15\% smaller and the estimated
  distance $0.5\kpc$ larger.  We consider this offset a systematic uncertainty.

Using our estimated distance, we calculate the source luminosity for each of the
pre-burst spectra reported in Table \ref{tab:preburst}. We find values ranging
from $1.4\E{37}\lumcgs$ to $3.5\E{37}\lumcgs$, which amounts to $4-10\%$ of
the Eddington luminosity.

\subsection{Burst phenomenology}
\label{sec:burst interpretation}

The X-ray bursts of \src show a rather interesting phenomenology. Based on the
light curves shown in Figure \ref{fig:burst lc}, we can think of the bright
bursts as a superposition of two components. Over the first five seconds, the
bursts show a ``fast'' component: the count-rate rises quickly, stabilizes for
a second or two, before dropping back down. At about 5 seconds after onset, the
burst profile becomes dominated by a second ``slow'' component, which causes
a shoulder in the light curve before continuing as a slower decay. The weak
bursts can then be interpreted as showing only the second ``slow''
component (see Figure \ref{fig:burst lc}).

  All X-ray bursts were detected while \src was in the soft state, hence the
  difference between the weak and bright types is not related to the accretion
  state in any obvious way. Even when accounting for variations of the
  intensity and hardness ratio within the soft state (Figure \ref{fig:light
  curve}), we find that either burst type can occur for the same
  conditions. The only evident difference is that the weak bursts tend to occur
  at higher (persistent) intensities, generally while the source flux is above
  $\approx5\E{-9} \fluxcgs$ (see Table \ref{tab:preburst}), or about 7.5\% of
  the Eddington luminosity. We caution against over-interpreting this finding,
  however, as the bright bursts are the more common burst type. Hence, the
  non-detection of weak bursts at lower intensities may simply be a sampling
  artifact. 

From our spectroscopic analysis, we found that the ``fast'' component of the
bright bursts is associated with a photospheric radius expansion phase. Indeed,
none of the weak bursts show evidence for PRE, although we note that many still
reach fluxes near the Eddington limit. Two additional features emerge when we
compare the time resolved spectroscopy of the different types of bursts. 
First, the time at which the PRE phase ends and the photosphere settles back on
the stellar surface coincides with the time at which the weak bursts reach
their peak flux. Second, the blackbody temperature shows the same time evolution
in all X-ray bursts.

It is tempting to associate the ``fast'' and ``slow'' components of the light
curve directly with the nuclear processes that power an X-ray burst. The
``fast'' component, with its fast evolution and high luminosity, has the
hallmarks of helium burning and, if seen in isolation, would be interpreted
as a pure helium burst \citep{Fujimoto1981,Narayan2003}. By comparison, the
``slow'' component could then be attributed to hydrogen burning via the rapid-proton (rp)
capture reaction chain \citep{Wallace1981, Schatz2001}, which tends
to proceed much more slowly. The difference between the weak and bright bursts
is one of ignition depth, with the bright bursts being due to ignition in a deep
layer of pure helium, while the weak bursts are due to ignition in a shallower
layer containing hydrogen. While this picture may explain some of the basic
systematics we observe, there are more subtle effects in our data that deserve
attention.

\subsubsection{Burst energy and recurrence}
First, let us point out that both bright and weak bursts have similar
timescales of about $10-15$\s. This is still much shorter than the $20\sim40\s$
typically seen for bursts with a high hydrogen abundance \citep[the most
notable example being GS 1826$-$24;][]{Galloway2008, Galloway2020}. Hence, if some
hydrogen is present in the burst fuel, its abundance is probably modest and not
too dissimilar between the two kinds of bursts. Optical observations of \src
have detected strong hydrogen and weak helium emission lines \citep{AtelStrader21},
indicating that the accreting matter is likely hydrogen rich. Given that the
burst profiles do not reflect this abundance, it seems likely that hydrogen
is burning stably between bursts through the hot CNO cycle. 

The duty cycle at which we sampled \src is on average about $4\%$. Given this
comparatively sparse coverage, we cannot reliably determine the waiting time
between individual bursts. As a first order estimate of the average burst recurrence,
however, we could assume that the bursts occurred at a roughly constant rate
and divide the total collected exposure by the number of observed bursts.
Taking the 2022 data, which contains all but one of our bursts, we have an
unfiltered exposure of 574\ks, giving $t_{\rm recurrence}
= 10_{-2}^{+3}\hr$, with the uncertainties following from Poisson counting
statistics. For comparison, the shortest waiting time between observed bursts
was $10.9\hr$ for bursts $\#12$ and $\#13$, suggesting that the average
recurrence is reasonably accurate at later times in the outburst. 

If we adopt the average recurrence time of $10$ hours, we can calculate the
implied $\alpha$ ratio for each observed burst as \citep{Gottwald1986,
Galloway2008}
\begin{equation} \label{eq:alpha}
  \alpha = \frac{F_{\rm preburst} \Delta t_{\rm recurrence}}{E_{\rm burst}},
\end{equation}
with the resulting values listed in Table \ref{tab:burst properties}. If we further
assume that each X-ray burst burns through all accreted matter, then the theoretically
expected $\alpha$ is $\approx40$ for hydrogen rich X-ray bursts and $100\sim150$
for pure helium bursts \citep{Galloway2008}. For some of the later bursts 
this assumed recurrence time is compatible with a burst that is mostly due
to helium ($\#12$ and up), as expected from the burst profiles. For the earlier
bursts, however, the estimated values for $\alpha$ are unreasonably high. In
particular, the weak bursts exceed nominal $\alpha$ values by a factor $4-5$,
while the bright bursts have a more modest factor $2-3$ excess. 
This suggests that either the recurrence time was much shorter for the earlier
X-ray bursts, that not all matter accreted between bursts contributes to the
burst fluence, or both. 

Traditional 1D theoretical models of burst ignition predict that the burst rate
should increase with mass accretion rate \citep{Fujimoto1981,Narayan2003}.
Hence, the higher accretion rate of the earlier bursts could well mean that
these bursts had recurrence times much shorter than our assumed $10$ hours. 
If we assume that bursts \#12 and \#13 are indeed consecutive bursts, then
their recurrence time gives us $\alpha=140\pm12$. By adopting this measured $\alpha$
for all bright bursts, we can use Eq. \ref{eq:alpha} to calculate the
implied recurrence times. We find that these recurrence times span
$5.6-14.4\hr$ (see Table \ref{tab:burst properties}).
There is an issue with the lower bound of these burst recurrence times, however: it
does not leave enough time for the CNO cycle to reduce the hydrogen fraction in the
accretion column. For solar abundances it takes approximately 10 hours for the
CNO cycle to deplete hydrogen at the base of the accretion column
\citep{Lampe2016}. Hence, if the implied recurrence times were correct, then
we would expect that the hydrogen abundance in these bursts to be changing
as well. Yet, such a change in abundance is not apparent in the burst profiles. 

In practice, there is another issue to consider: the relation between the mass
accretion rate and the burst recurrence does not always hold. At sufficiently
high mass accretion rate the burst rate is observed to decrease
\citep{Cornelisse2003, Galloway2008}. Additionally, the critical accretion rate
at which that turn over occurs is a function of the stellar spin frequency
\citep{Galloway2018, Cavecchi2020}. For \src, with a presumed spin frequency of
$585\hz$ \citep{Li2022}, the burst rate likely peaks around $5\%$ Eddington
\citep{Cavecchi2020}. Hence, we would expect the burst rate to decrease as the
mass accretion rate goes up, which is opposite to the trend needed to explain
our results. 

One way our data might be reconciled with theory is if the $585\hz$ burst
oscillation reported by \citet{Li2022} does not correspond to the neutron star
spin frequency. While the high signal strength of this oscillation makes it
unlikely that it was a spurious detection, we note that the oscillation was
only significant for about $1\s$. This is a much shorter signal duration
than normally seen for burst oscillations \citep{Watts2012, Bilous2019}, even
in the NICER band \citep{Bult2019}. Hence, there remains a 
possibility that the oscillation reported by \citet{Li2022} is what
\citet{Bilous2019} called a ``glimmer'' and that it is unrelated to the neutron
star spin frequency. 

\subsubsection{Ignition latitude}
The more likely interpretation of our data is that not all accreted
matter burns in the X-ray burst. If some part of the accreted envelope
is not involved in the X-ray burst, then the $\alpha$ estimated through
Equation \ref{eq:alpha} will overestimate the $\alpha$ predicted from theory.
One way this might happen is if the formation of a boundary layer concentrates
accretion onto the stellar surface in the equatorial region
\citep{Inogamov1999, Spitkovsky2002}. The local accretion rate at the equator
could then be high enough to support stable burning of both hydrogen and
helium, while still allowing unstable helium ignition at higher latitudes
\citep{Cavecchi2017}.

There are several features in \src that appear to support the idea of a
boundary layer causing equatorial stable burning. 
The first is the fact that the weak bursts tend to have slow
rise with a tendency toward a concave shape. Such a concave burst rise has been
argued to be a signature of burst ignition near the stellar poles
\citep{Maurer2008}. Because both the ignition conditions and the flame
spreading speed should depend on latitude \citep{Spitkovsky2002, Cooper2007a},
this could then also naturally explain the difference between weak and bright
bursts, with the former originating near the poles and the latter occurring at
less extreme latitudes. 
A second relevant observation is that \src is a prominent source of mHz QPOs
(Mancuso et al. 2022, in prep). Such mHz QPOs are generally believed to be due to
marginally stable burning \citep{Heger2007, Altamirano2008c}, and may arise
naturally at the boundaries of a stable burning region around the equator
\citep{Cavecchi2020}.
Finally, a third effect comes from the excess flux in the burst spectrum, which
again points to the presence of a boundary layer. We discuss this point in
Section \ref{sec:nonthermal}.

\subsubsection{Short recurrence bursts}
Finally, as an alternative explanation for the weak bursts, we consider the
possibility that they are short time recurrence bursts: X-ray bursts that occur
within a few minutes to at most half an hour after the previous event
\citep{Boirin2007, Keek2010}. Such short recurrence bursts are not due to the
accumulation of a fresh accretion column, but instead thought to be caused by
the turbulent mixing bringing unburned material from the previous burst down to
ignition depth. Observationally, such events are often observed as a train of
X-ray bursts that tend to become progressively weaker. Hence, the weak
bursts of \src could be interpreted as one of the later events of such a short
recurrence train. 
While we did not observe any such short recurrence burst trains, the low duty
cycle of our sampling means that we cannot rule them out either. Typically, our observations
span only about $\sim1000\s$ around an observed X-ray burst, leaving room for recurrences
of $\gtrsim10$ minutes.  
More importantly, however, \citet{Keek2010} found that short recurrence bursts
tend to be cooler than the primary burst, and lack the slower decay component
due to the rp-process. This phenomenon is believed to be due to the fact that
these short recurrence events occur in a hydrogen depleted environment. The
weak bursts of \src do not exhibit this behavior. While fluence of the weak
bursts is lower, their peak flux, temperature evolution, and duration is more
or less the same as for the bright bursts. Hence, the weak bursts are likely
not short recurrence time events.

\subsection{The enhanced burst emission}
\label{sec:nonthermal}

All X-ray bursts observed from \src showed a significant departure from a pure
thermal spectrum. We accounted for the excess flux in X-ray burst emission by
using the $f_a$ model of \citet{Worpel2013}. The measured values of $f_a$ were
found to roughly correlate with the burst intensity. The weak bursts had peak $f_a$ values
of about $2-3$, while the bright PRE bursts showed peak $f_a$ values between
$7-10$. These results are similar to those found with RXTE \citep{Worpel2013,
Worpel2015} as well as more recent studies with NICER \citep{Guver2022a,
Guver2022b}. A more subtle effect apparent in our observations is that the
X-ray bursts observed at later times appear to have higher peak $f_a$ values.
That is, PRE bursts observed when the persistent intensity is high tend to peak
at values of about $7$, while the later time X-ray bursts (notably, $\#16$ and
$\#17$) reach much higher values of about $10$. This pattern suggests that the
magnitude of the flux excess depends not only on the intensity of the
thermal burst emission, but also on the mass accretion rate (or perhaps more
appropriately, on the accretion state). 

The detection of excess burst emission has become increasingly
common in recent year, either through the increased sensitivity afforded by
NICER \citep{Keek2018a,Keek2018b, Bult2019, Jaisawal2019, Guver2022a,
Guver2022b}, or through broadband X-ray coverage with AstroSat
\citep{Bhattacharyya2018, Roy2021, Kashyap2022} or Insight-HXMT
\citep{Chen2019}. While the emission of a neutron star atmosphere
is expected to deviate from Plank's law \citep{London1986, Madej2004, Suleimanov2011},
the observed deviations from a blackbody spectrum are well in excess of
what an atmosphere model can explain \citep{Zand2017}. Instead, this flux
excess is understood to be an indicator of interactions between the burst
emission and the accretion flow surrounding the neutron star
\citep{Degenaar2018}, even if the precise nature of that interaction remains
a topic of investigation. Of the various interactions that can take place,
three processes are commonly considered to explain the excess flux.
First, the burst emission is reprocessed in the surface layers of the accretion
disk and ``reflected'' back into the line of sight \citep{Ballantyne2004a, Ballantyne2004b}
thereby adding a secondary reflection component to the spectrum, whose
magnitude should approximately follow the burst intensity. 
Second, the burst radiation could induce a Poynting-Robertson (PR) drag in the inner
accretion disk \citep{Walker1989, Walker1992}, which temporarily increased the
accretion rate onto the stellar surface. 
Third, the stellar surface could be covered by a boundary layer,
such that our view of the neutron star is (partially) obscured by a
Comptonizing medium \citep{Kajava2014, Koljonen2016}. 

Given the uncertainty about its origin, the presence of a flux excess makes
it more challenging to discern how much of the observed X-ray emission is
directly due to the thermonuclear processes in the stellar envelope, and how
much is added by the interaction with the accretion disk. In our analysis, we
therefore calculated both the total flux in the X-ray burst (minus the pre-burst contribution)
and the blackbody contribution separately. As shown in Figure \ref{fig:flux
temp}, we found that the flux of the blackbody component decreases during the
PRE phase, while the total burst flux remains constant. We take this to mean
that it is the total burst flux that corresponds to the Eddington luminosity,
which implies that the deviation away from pure blackbody emission is due to a
(Comptonizing) scattering medium that (partially) obscures the line of sight to the neutron
star. This behavior is naturally explained by the boundary layer
interpretation. The disk reflection and PR drag mechanisms cannot be entirely
ruled out, though. Numerical simulations of the disk response to the X-ray
burst radiation suggests that both reflection and PR drag can occur
simultaneously, and act to increase the scale height of the X-ray burst
\citep{Fragile2020}. Depending on the resulting height of the disk and the
binary inclination of \src, it is possible that the inflated disk temporarily
obscures the neutron star. 

\begin{figure}[t]
  \includegraphics[width=\linewidth]{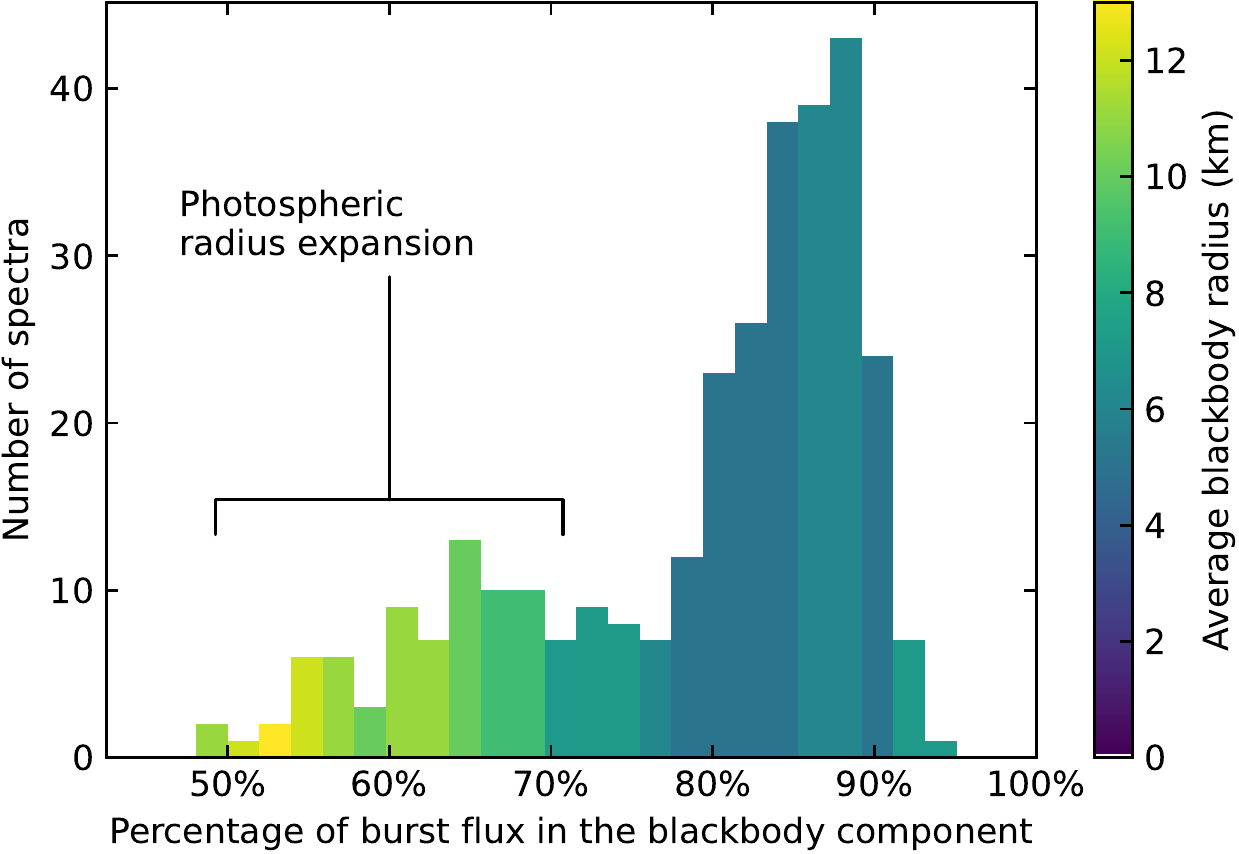}
  \caption{%
    Histogram of the flux fraction contributed by the blackbody component.
    Each histogram bar is color coded according to the average blackbody
    normalization in that bin. 
  } 
  \label{fig:flux distribution}
\end{figure}

In this context, it is worth asking how much of the total observed flux is
contributed by either the blackbody or the non-thermal component. In Figure
\ref{fig:flux distribution} we show a histogram of X-ray burst spectra as
a function of the blackbody contribution. The distribution clearly has two
components, with a primary peak at $\approx85\%$ and a secondary component at $\approx65\%$. By
calculating the average blackbody normalization of each histogram bin, we
further see that secondary component is associated with the PRE phase of bursts.
Thus, during PRE the non-thermal emission accounts for some $\approx35\%$ of
the total flux, while outside of the PRE phase, this contribution reduces to
$\approx15\%$.  
That the non-thermal flux contribution increases during the PRE phase suggests
that whatever the scattering medium is, be it the disk or the boundary layer,
it must be evolving under the influence of sustained irradiation by the burst
emission. If so, we would expect the shape of the non-thermal emission to change
over the course of an X-ray burst. 

\subsection{A relation with double peaked bursts}
The shape of the bright bursts observed from \src invites a comparison to
double peaked X-ray bursts.  Such double peaked bursts are characterized by
a secondary peak in the bolometric X-ray flux that is astrophysical in origin
(meaning that it is not due to the blackbody spectrum shifting out of the
instrument passband).  A number of models have been proposed to explain double
peaked X-ray bursts, including waiting points in the rp-process
\citep{Fisker2004, Fisker2008}, stalled flame spreading \citep{Bhattacharyya2006}, remixing
of unburned material \citep{Keek2017}, and line of sight absorption
\citep{Kajava2017}. 

Although the bursts of \src do not show a secondary increase in bolometric
flux, they do show a shoulder in their flux evolution. We can speculate that
the process causing the flux to temporarily form a plateau may be (in part) the
same process behind the second peak in other bursters.  Indeed, the time
resolved spectroscopy of the bright bursts from \src appears to be very similar
to those of a double peaked burst from 4U 1608$-$52 observed with NICER
\citep{Jaisawal2019}, lending some support to this assumption. 

Phenomenologically, one can imagine that the difference between observing a plateau
and a secondary flux increase is simply one of the relative timescales for the
PRE phase (what we called the ``fast'' component) and the rise time of the weak
bursts (what we called the ``slow'' component). That is, if the PRE phase is
faster than the slow rise, one sees a clear secondary peak \citep[see
e.g.,][]{Li2021}. If these two timescales are roughly similar (as is the case
here), one sees a shoulder in the light curve. Finally, when the PRE phase
lasts longer than the slow rise, one might see a break in the burst tail (e.g.
burst $\#10$). If these phenomena are indeed related, that would support the
rp-process waiting point model, as it suggests that the secondary peaks are at
least partially due to a separation of the hydrogen and helium burning
processes.

In this context, one might also ask if other X-ray bursters exhibit behavior
similar to what we observe in \src. Indeed, the type of shoulders we find in the
bright bursts are also seen in the profiles of prominent bursters such as
4U 1636$-$536 and Aql X-1 \citep{Galloway2008, Guver2022a, Guver2022b} and have
been previously attributed to the rp-process \citep{Zand2017}. Interestingly,
these two bursters also show weak bursts that appear to have a similar (if much less
pronounced) alignment to cooling tails of their brighter counter-parts
\citep[see, e.g.,][]{Guver2022b}. Additional similarities between these sources
exist: like \src, both 4U 1636-536 and Aquila X-1 have high ($>500$ Hz) spin
frequencies \citep{Strohmayer1998, Zhang1998, Casella2008} and are known to
show mHz QPOs \citep{Revnivtsev2001}. On the other hand, Swift J1858.6$–$0814 is
another example of an X-ray burster that shows both mHz QPOs and very similar
(bright) X-ray burst profiles \citep{Buisson2020}, yet there the shoulder
appears at different intensities for different X-ray bursts. Hence, physical
processes beyond what we considered in this paper may be in play. A  detailed
comparison of these various sources is beyond the scope of this work, but may
make for interesting topic of future investigation.

\section{Conclusion}
We have presented a spectroscopic analysis of 17 thermonuclear X-ray
bursts observed from the neutron star X-ray transient \src. We observed
photospheric radius expansion in twelve of these X-ray bursts, allowing us to
estimate the source distance at $6.9\pm0.2\kpc$ with an additional systematic
uncertainty of $0.5\kpc$. 

We found that \src shows both weak and bright bursts, and that these two types
have a striking visual alignment: the bright bursts show a shoulder in their
profile at the time that the weak burst reach peak intensity.  We suggest that
this alignment points to two mostly independent nuclear burning processes:
helium burning powering the bright photospheric radius expansion phase, and
rp-capture hydrogen burning setting the slower (and weaker) cooling tails.

Although multiple interpretations might explain why both bright and weak bursts
are observed from this source, we found that the burst properties could be
naturally explained if we assume that accretion onto the neutron star proceeds
through a boundary layer, such that accreted matter burns stably at the stellar
equator, while unstable ignition occurs at higher latitudes.

\nolinenumbers
\facilities{ADS, HEASARC, NICER}
\software{heasoft (v6.30), nicerdas (v9)}

\begin{acknowledgments}
  \nolinenumbers
  This work made use of data and software provided by the High Energy
  Astrophysics Science Archive Research Center (HEASARC). 
  P.B. acknowledges support from NASA through the Astrophysics Data Analysis
  Program (80NSSC20K0288) and the CRESST II cooperative agreement (80GSFC21M0002).
  G.C.M. was partially supported by Proyecto de Investigaci\'on Plurianual
  (PIP) 0113 (Nacional de Investigaciones Cient\'{\i}ficas y T\'ecnicas
  (CONICET)) and from PICT-2017-2865 (Agencia Nacional de Promoci\'on
  Cient\'ifica y Tecnol\'ogica (ANPCyT)).
  D.A. acknowledges support from the Royal Society.
  S.G. acknowledges the support of the CNES.
\end{acknowledgments}


\end{document}